\documentclass[journal,twoside]{IEEEtran}
\renewcommand{\IEEEQED}{}

\ifCLASSINFOpdf
\else
\fi
\usepackage{xfrac}
\usepackage{amsmath}
\usepackage{multirow}
\usepackage{color}
\usepackage{amssymb}
\usepackage{cite}

\usepackage{stfloats}

\begin{document}
\bstctlcite{MyBSTcontrol}
\title{Error Probability Analysis of NOMA-based Diamond Relaying Network}
%
%
%
\author{Ferdi Kara, ~\IEEEmembership{Member,~IEEE,}
        Hakan Kaya 
\thanks{Copyright (c) 2015 IEEE. Personal use of this material is permitted. However, permission to use this material for any other purposes must be obtained from the IEEE by sending a request to pubs-permissions@ieee.org.}
\thanks{This work is supported by TUBITAK under 2211-E program and by Zonguldak Bulent Ecevit University with grant no: 2019-75737790-01}
\thanks{The authors are with EEE department at Zonguldak Bulent Ecevit University, Incivez, Zonguldak 67100, Turkey e-mail:\{f.kara,hakan.kaya\}@beun.edu.tr }}
%
%
\markboth{IEEE Transactions on Vehicular Technology,~Vol.~xx, No.~xx, Month~2019}%
{Kara and Kaya: Error Probability Analysis of NOMA-based Diamond Relaying Network}
%
\maketitle
\begin{abstract}
Non-orthogonal multiple access (NOMA)-based cooperative relaying systems (CRS) are very promising to overcome spectral inefficiency of conventional cooperative communications. Although NOMA-CRS have great recent attention, almost all studies investigate NOMA-CRS only in terms of capacity and outage probability. Error performances of NOMA-CRS have not been well-studied. In this paper, we analyze error performance of NOMA-based diamond relaying network (NOMA-DRN) with imperfect successive interference canceler (SIC) as a NOMA-CRS scheme. We derive exact bit error probability (BEP) for NOMA-DRN and provide a tight approximated BEP in the closed-form. In addition, high-SNR analysis is conducted to present that NOMA-DRN has an error floor. Moreover, it is proved that NOMA-DRN turns out to be a non-equiprobable communication system and we derive priori probabilities of symbols. All derived expressions are validated via computer simulations.
\end{abstract}

\IEEEpeerreviewmaketitle
\section{Introduction}
\IEEEPARstart{N}{on}-orthogonal multiple access (NOMA) is seen as one of the key technologies to fulfill requirements of future wireless networks \cite{Poor2015}. NOMA is based on allowing users to share the same resource blocks (time, frequency, code) by splitting them into power domain. It provides a spectral efficient communication and makes the massive connection possible in such networks, Internet of Things (IoT). Thus, a two-users application of NOMA has already taken place in the wireless standards \cite{PP2016} and it is expected to be part of wireless evolution. In addition to its spectral efficiency, since it is easy to implement in other physical layer techniques  (e.g., cooperative communication, MIMO, cognitive radio etc.), NOMA involved systems have taken tremendous recent attention \cite{Vaezi2019}.

The interplay between NOMA and cooperative communication has been one of the most attracted topics to increase spectral efficiency of cooperative communications. The authors in \cite{Kim2015a} have firstly proposed NOMA-based cooperative relaying systems (CRS). Ergodic sum rate of NOMA-CRS has been analyzed and its superiority to conventional CRS is proved over Rayleigh \cite{Kim2015a} and Rician fading channels \cite{Jiao2017}. The authors in \cite{Zhang2018} have analyzed the outage probability of NOMA-CRS for various scenarios with imperfect channel state information (CSI). Then, NOMA-CRS with an amplify-forward relay rather than decode-forward relay has been proposed in \cite{Abbasi2019}, and approximate analysis is given for achievable rate. Moreover, according to relay positions or the presence of direct link between source and destination, various NOMA-CRS have been considered in literature. In one of them, NOMA-based diamond relaying has been proposed where two relays are located between source and destination with different distances and its achievable rate has been investigated \cite{Wan2019}. However, the aforementioned studies are only devoted to analyze two of the key performance indicators (KPIs) (i.e., sum-rate and outage) and assume that successive interference canceler (SIC) is perfect, thus the effects of errors during SIC have not been presented. Although error performance is one of the most important KPIs, to the best of authors' knowledge, error performance of NOMA-CRS has only been analyzed in \cite{Li2019} for multiple antenna situations and approximate bit error probability (BEP) expressions are derived, yet exact expressions have not been provided.

In this paper, we investigate error performance of NOMA-based diamond relaying network (NOMA-DRN) as a subset of NOMA-CRS and derive end-to-end (e2e) BEP to evaluate symbol level error performance in the presence of SIC errors. The trade-off between error performance and achievable rate for NOMA-DRN is raised. The remainder of the paper is as follows. In Section II, we introduce the NOMA-DRN. Then, in Section III, analytical analysis for e2e BEP is provided. Validation of the derived expressions via Monte Carlo simulations are presented in Section IV. Finally in Section V, results are discussed and the paper is concluded.
\section{System Model}
We consider a cooperative relaying system where two relays (i.e., $R_k, k=1,2$) are located between source (S) and destination (D). The distances of each relay to source and to destination differ, hence the system model is named as diamond relaying as in \cite{Wan2019}. We assume that all nodes are equipped with single antenna and the channel coefficient between each node follows $CN(0,\sigma^2_\lambda)$ $i.e., \lambda=SR_1,SR_2, R_1D, R_2D$. $\sigma^2_\lambda=\mu d_\lambda^\tau$ where $d_\lambda$ is the distance between nodes and we assume that $d_{SR_1}>d_{SR_2}$ while $d_{R_1D}<d_{R_2D}$. $\mu$ and $\tau$ denote the propagation constant and path-loss exponent, respectively. Relays operate in half-duplex mode and total communication is completed in two phases (time slots). In the first phase of communication, source implements superposition-coding for two symbols and transmits them simultaneously on the same resource block (downlink-NOMA). The received signals by the relays are given as
\begin{equation}
y_\lambda=\sqrt{P_S}\left(\sqrt{\alpha_1}x_1+\sqrt{\alpha_2}x_2\right)h_\lambda+n_\lambda, \ \lambda=SR_1, SR_2,
\end{equation}
where $P_S$ is the transmit power of source. $\alpha_1$ and $\alpha_2$ are the power allocation coefficients for the base-band symbols of $x_1$ and $x_2$ (modulated by BPSK\footnote{BPSK is chosen since error analysis for cooperative communication is mostly provided for BPSK in literature.}), respectively. $\alpha_1>\alpha_2$ and $\alpha_1+\alpha_2=1$. $h_\lambda$ is the channel coefficient and $n_\lambda$ is additive Gaussian noise which follows $CN(0,N_0)$. After receiving signals, $R_1$ detects $x_1$ symbols by pretending $x_2$ symbols as noise whereas $R_2$ should implement SIC to detect $x_2$ symbols. Then, in the second phase of communication, each relay forwards detected symbols to the destination (uplink-NOMA) and the received signal at the destination is given as
\begin{equation}
y_D=\sqrt{P_R}\left(\sqrt{\beta_1}\hat{x}_1h_{R_1D}+\sqrt{\beta_2}\hat{x}_2h_{R_2D}\right)+n_D,
\end{equation}
where $P_R$ is the total power of relays and it is allocated\footnote{Although the relay nodes could have independent power constraints, for the total power consumption such assumption is reasonable and has been made in existing studies \cite{Wan2019}.} by $\beta_1$ and $\beta_2$ where $\beta_1+\beta_2=1$. $\hat{x}_1$ and $\hat{x}_2$ are the detected/estimated symbols at the relays. Finally, the destination firstly detects $x_1$ symbols and subtracts it from total received signal (i.e., SIC) and detects $x_2$ symbols.
The system model is given in Fig. 1.
\begin{figure}[!t]
    \centering
    \includegraphics[width=8cm,height=3cm]{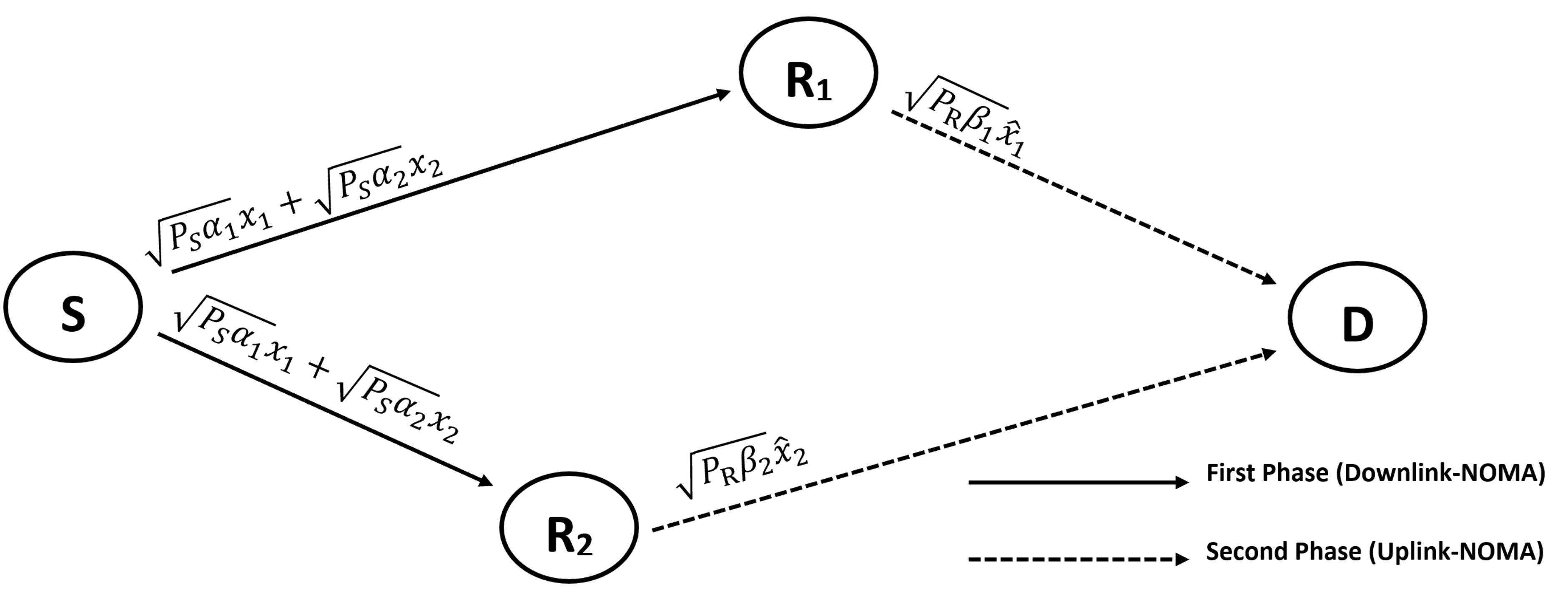}
   \caption{The illustration of NOMA-DRN}
    \label{system_model}
\end{figure}
\section{Performance Analysis}
\subsection{Error Analysis of First Phase (Downlink-NOMA)}
Since two symbols are transmitted on the same resource block, the symbols are exposed to inter-symbol-interferences. The total received symbols are represented\footnote{For simplicity, the effects of channel fading and noise are not represented.} in Fig. 2.a\textit{(i)}. In Fig 2.a, $\epsilon_i$ denotes the energy levels of symbols after superposition coding or SIC. One can easily see that the received signals have different energy levels according to which $x_1$ and $x_2$ symbols are sent (i.e., events A or B). Thus, in error analysis, we should also consider two different cases with their priori probabilities \cite{Kara2018d}.
We firstly derive the BEP for $x_1$ symbols at $R_1$ (i.e., $P_I^{(SR_1)}$).
$R_1$ implements maximum-likelihood (ML) detector to detect $x_1$ symbols, hence ML decision rule for BPSK (i.e., whether $y_{SR_1}<0$ or $y_{SR_1}\geq0$) should be considered. We firstly consider the left side (LS) of received signal at the $R_1$ of Fig. 2.a\textit{(i)}, hence the error probability of erroneous detection is determined as
\begin{figure}[!t]
    \centering
    \includegraphics[width=8cm,height=7.5cm]{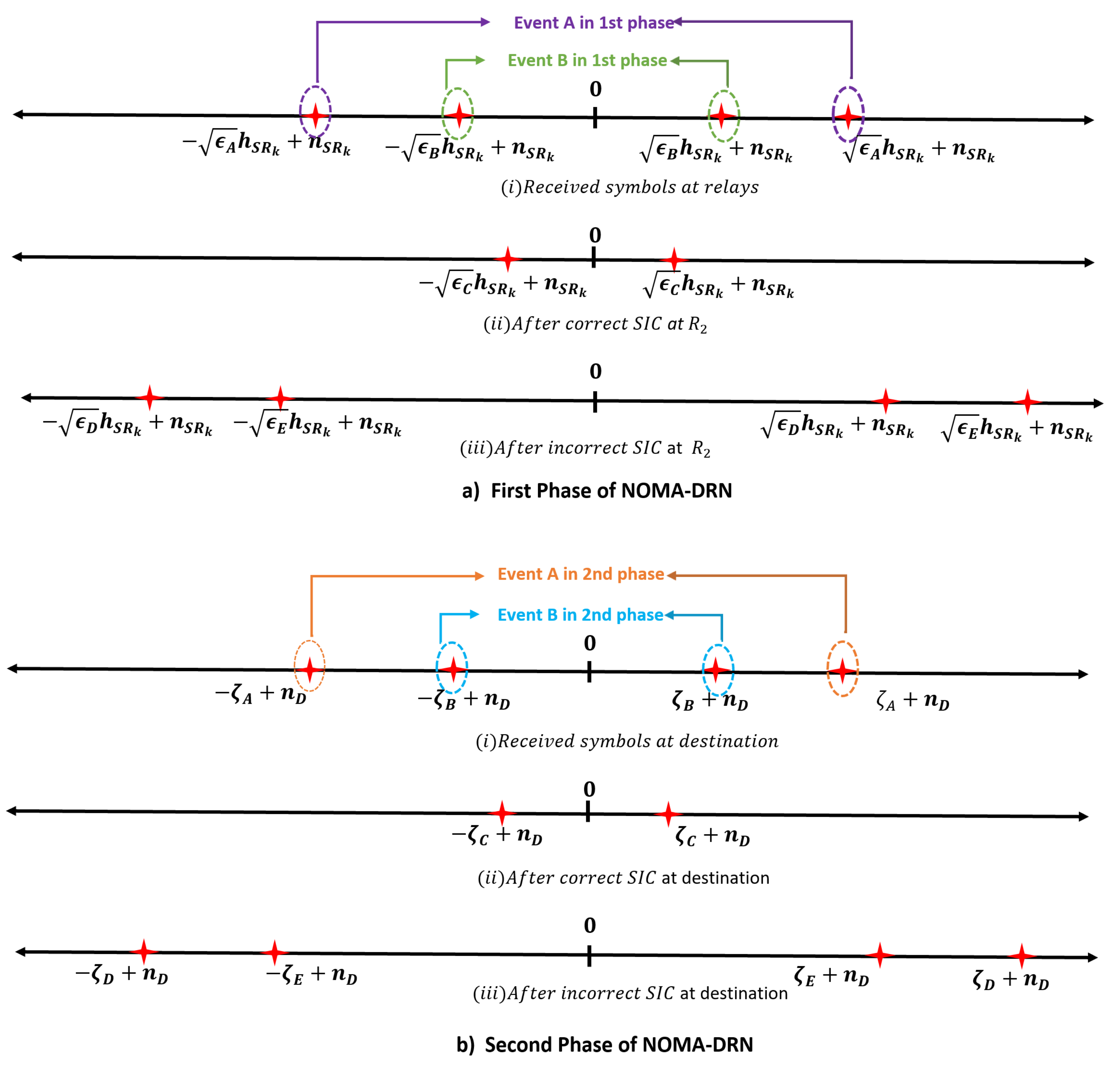}
    \caption{Illustration of received signals a) First Phase (Downlink-NOMA) b) Second Phase (Uplink-NOMA)}
    \label{fig1}
\end{figure}
\begin{equation}
\begin{split}
&P_I^{(SR_1)}(e|_{h_{SR_1}})= \sum_i p(i_{1st})P_I^{(SR_1)}(e|_{h_{SR_1}\cap i})  \\
&=\sum_i p(i_{1st}) P(n\geq \sqrt{\epsilon_i}\left|h_{SR_1}\right|), \ \ i=A,B,
\end{split}
\end{equation}
where  $p(i_{1st}) \quad i=A,B$ denotes the priori probability of event $i$ may occur in the first phase and $\epsilon_i$ is the energy of the total superposition-coded symbol in the event $i$. $\epsilon_A=\left(\sqrt{\alpha_1P_S}+\sqrt{\alpha_2P_S}\right)^2$, $\epsilon_B=\left(\sqrt{\alpha_1P_S}-\sqrt{\alpha_2P_S}\right)^2$. $\epsilon_A$ and $\epsilon_B$ are obtained by utilizing $x_1=\mp1$ and $x_2=\mp1$ for BPSK. It is noteworthy that noise term (i.e., $n$) in (3) is the in-phase component of AWGN with $\sfrac{N_0}{2}$ variance since only the in-phase component of noise has an effect on decision. The error probability for right side (RS) of received signal at the relay becomes same with (3). Thus, conditional error probability of $x_1$ at $R_1$ can be easily determined as
\begin{equation}
P_I^{(SR_1)}(e|_{h_{SR_1}})= \sum_i p(i_{1st}) Q\left(\sqrt{2\epsilon_i\gamma_{SR_1}}\right), \ i=A,B, \\
\end{equation}
where $\gamma_{\lambda}=\sfrac{\left|h_{\lambda}\right|^2}{N_0}$. We assume that $x_1$ and $x_2$ symbols have equal probability (i.e., $p(A_{1st})=p(B_{1st})=0.5$) and considering the $\gamma_{SR_1}$ follows exponential distribution, by averaging (4), the average BEP (ABEP) of $x_1$ symbols at $R_1$ is determined as
\begin{equation}
P_I^{(SR_1)}(e)= \sum_i\frac{1}{4}\left(1-\sqrt{\frac{\epsilon_i\sigma_{SR_1}^2}{N_0+\epsilon_i\sigma_{SR_1}^2}}\right), \ i=A,B.
\end{equation}

On the other hand, in order to derive error probability of $x_2$ symbols at $R_2$ (i.e., $P_{II}^{(SR_2)}$), we should consider two cases whether $x_1$ symbols are detected correctly (correct SIC) or erroneously (erroneous SIC). Firstly, we assume that $x_1$ symbols are detected correctly. In this case, after SIC, only $x_2$ symbols with $h_{SR_2}$ coefficient and the noise $n_{SR_2}$ remain as in Fig. 2.a\textit{(ii)}. However, we hereby note that this has conditional probability of correct detection of $x_1$ at $R_2$, hence with the aid of (3), it is derived as,
\begin{equation}
\begin{split}
&P_{II}^{(SR_2)}(e|_{h_{SR_2}\cap correct_{x_1}})=p(A_{1st}) P({n<\sqrt{\epsilon_A}\left|h_{SR_2}\right|}) \\
 &P(n\geq \sqrt{\epsilon_C}\left|h_{SR_2}\right| \bigm|{n<\sqrt{\epsilon_A}\left|h_{SR_2}\right|}) \\
&+ p(B_{1st}) P({n<\sqrt{\epsilon_B}\left|h_{SR_2}\right|}) \\
 &P(n<-\sqrt{\epsilon_C}\left|h_{SR_2}\right| \bigm|{n<\sqrt{\epsilon_B}\left|h_{SR_2}\right|}), \\
\end{split}
\end{equation}
where $\epsilon_C=\alpha_2P_S$. In (6), $P({n<\sqrt{\epsilon_i}\left|h_{SR_2}\right|}), i=A,B$ denotes the priori probability of correct detection for $x_1$ symbols at $R_2$ which is easily derived with the aid of (3) by considering correct decision and changing channel coefficient. Conditional probabilities denote the error probability for remained signal under the condition $x_1$ symbols have been already detected correctly. Recalling that a conditional probability of events $P(\varphi\bigm|\theta)=\sfrac{P(\varphi\cap\theta)}{P(\theta)}$ is defined. Thus, after some algebraic manipulations, the error probability of $x_2$ at $R_2$, in case $x_1$ symbols are detected correctly, is determined as
\begin{equation}
\begin{split}
&P_{II}^{(SR_2)}(e|_{h_{SR_2}\cap correct_{x_1}})=p(A_{1st})\left(Q\left(\sqrt{2\epsilon_C\gamma_{SR_2}}\right) \right.\\
&\left.-Q\left(\sqrt{2\epsilon_A\gamma_{SR_2}}\right)\right)+p(B_{1st}) Q\left(\sqrt{2\epsilon_C\gamma_{SR_2}}\right).\\
\end{split}
\end{equation}
Then, we assume that $x_1$ symbols are detected erroneously at $R_2$. In this case, the remained signal after subtracting erroneous estimated $\hat{x}_1$ from received signal is given in Fig. 2.a\textit{(iii)}. Similar to (6), considering wrong detection condition, the conditional error probability including the priori probability of error detection for $x_1$ symbols is determined as
\begin{equation}
\begin{split}
&P_{II}^{(SR_2)}(e|_{h_{SR_2}\cap error_{x_1}})=p(A_{1st}) P({n\geq\sqrt{\epsilon_A}\left|h_{SR_2}\right|}) \\
 &P(n\geq \sqrt{\epsilon_D}\left|h_{SR_2}\right| \bigm|{n\geq\sqrt{\epsilon_A}\left|h_{SR_2}\right|}) \\
&+ p(B_{1st}) P({n\geq\sqrt{\epsilon_B}\left|h_{SR_2}\right|}) \\
 &P(n<\sqrt{\epsilon_E}\left|h_{SR_2}\right| \bigm|{n<\sqrt{\epsilon_B}\left|h_{SR_2}\right|}), \\
\end{split}
\end{equation}
where $\epsilon_D=\left(2\sqrt{\alpha_1P_S}+\sqrt{\alpha_2P_S}\right)^2$ and $\epsilon_E=\left(2\sqrt{\alpha_1P_S}-\sqrt{\alpha_2P_S}\right)^2$.
Then, after some algebraic manipulations, it is derived as
\begin{equation}
\begin{split}
&P_{II}^{(SR_2)}(e|_{h_{SR_2}\cap error_{x_1}})=p(A_{1st})Q\left(\sqrt{2\epsilon_D\gamma_{SR_2}}\right) \\
&+p(B_{1st}) \left(Q\left(\sqrt{2\epsilon_B\gamma_{SR_2}}\right)- Q\left(\sqrt{2\epsilon_E\gamma_{SR_2}}\right) \right).\\
\end{split}
\end{equation}
The BEP of $x_2$ symbols at $R_2$ is obtained by $P_{II}^{(SR_2)}(e|_{h_{SR_2}})=P_{II}^{(SR_2)}(e|_{h_{SR_2}\cap correct_{x_1}})+P_{II}^{(SR_2)}(e|_{h_{SR_2}\cap error_{x_1}})$. By averaging over instantaneous channel conditions, the ABEP for $x_2$ at $R_2$ is derived as
\begin{equation}
\begin{split}
&P_{II}^{(SR_2)}(e)=\frac{1}{2}(1-\sqrt{\frac{\epsilon_C\sigma_{SR_2}^2}{N_0+\epsilon_C\sigma_{SR_2}^2}})+\frac{1}{4}\left(\sqrt{\frac{\epsilon_A\sigma_{SR_2}^2}{N_0+\epsilon_A\sigma_{SR_2}^2}} \right.\\
&-\left.\sqrt{\frac{\epsilon_B\sigma_{SR_2}^2}{N_0+\epsilon_B\sigma_{SR_2}^2}}-\sqrt{\frac{\epsilon_D\sigma_{SR_2}^2}{N_0+\epsilon_D\sigma_{SR_2}^2}}+\sqrt{\frac{\epsilon_E\sigma_{SR_2}^2}{N_0+\epsilon_E\sigma_{SR_2}^2}}\right).
\end{split}
\end{equation}
\subsection{Error Analysis of Second Phase (Uplink-NOMA)}
\subsubsection{Exact Expressions for Second Phase}
Error analysis for uplink-NOMA can be determined as in donwlink-NOMA. However, it is noteworthy that the received symbols at destination are the combination of different symbols with different channel fading coefficients (i.e., $h_{R_1D}$ and $h_{R_2D}$) as given in Fig. 2.b\textit{(i)}. In Fig. 2.b, $\zeta_i$ denotes the superimposed symbol and the remained symbols after SIC including channel fading coefficients at the destination. Hence, the analysis should be modified according to these superimposed symbols. Since destination firstly decodes $x_1$ symbols by pretending $x_2$ symbols as noise, the  BEP is obtained by applying the steps (3)-(4) as
\begin{equation}
P_I^{(R_1D)}(e|_{\zeta_i})= \sum_i p(i_{2nd}) Q\left(\sqrt{2}\zeta_i\right), \ \ i=A,B, \\
\end{equation}
where $\zeta_A=\sfrac{\left(\sqrt{\beta_1P_R}\left|h_{R_1D}\right|+\sqrt{\beta_2P_R}\left|h_{R_2D}\right|\right)}{\sqrt{N_0}}$ and $\zeta_B=\sfrac{\left(\sqrt{\beta_1P_R}\left|h_{R_1D}\right|-\sqrt{\beta_2P_R}\left|h_{R_2D}\right|\right)}{\sqrt{N_0}}$. In order to obtain ABEP for $x_1$ symbols, we should average (11) over instantaneous channel fading. Hence, the joint probability density functions (PDF) should be derived firstly. The PDFs for $Z=X+Y$ and $W=X-Y$ are given as follow in \cite{Archer1967} and \cite[Appendix]{Kara2018d}, respectively, where $X$ and $Y$ are independent Rayleigh random variables with the scale parameters $\sigma_X$ and $\sigma_Y$, respectively.
\begin{equation}
\begin{split}
&p_Z\left(z\right)=
\\&
\frac{\sigma_X^2z}{\left({\sigma_X^2}+{\sigma_Y^2}\right)^2}exp\left(-\frac{z^2}{{2\sigma_X^2}}\right)+\frac{{\sigma_Y^2}z}{\left({\sigma_X^2}+{\sigma_Y^2}\right)^2}exp\left(-\frac{z^2}{{2\sigma_Y^2}}\right)
\\&
+\sqrt{\frac{\pi}{2}}\frac{\sigma_X\sigma_Y\left[z^2-\left({\sigma_X^2}+{\sigma_Y^2}\right)\right]}{\left({\sigma_X^2}+{\sigma_Y^2}\right)^\frac{5}{2}}exp\left(-\frac{z^2}{2\left({\sigma_X^2}+{\sigma_Y^2}\right)}\right)
\\&
\left[erf\left(\frac{z\sigma_Y}{\sigma_X\sqrt{2\left({\sigma_X^2}+{\sigma_Y^2}\right)}}\right)+erf\left(\frac{z\sigma_X}{\sigma_Y\sqrt{2\left({\sigma_X^2}+{\sigma_Y^2}\right)}}\right)\right]
\end{split}
\end{equation}
and
$p_W\left(w\right)=$
\begin{equation}
\begin{cases}
\frac{4e^{-\sfrac{w^2}{\sigma_X^2}}}{\sigma_X^2\sigma_Y^2}\left[\frac{\sqrt\pi erfc\left(\frac{w\left(1-\tau\sigma_X^2\right)}{\sqrt\tau\sigma_X^2}\right)\exp\left(\frac{w^2}{\tau\sigma_X^4}\right)}{2\tau^\frac{3}{2}\sigma_X^2}\times
 \right. \\  \left.
\left\{\frac{\tau\sigma_X^4+2w^2}{2\tau\sigma_X^2}-w^2\right\}-\frac{w\ exp\left(-w^2\left(\tau-\frac{2}{\sigma_X^2}\right)\right)}{2\tau^2\sigma_X^2}\right], & \text{$w<0$}.\\

\frac{4e^{-\sfrac{w^2}{\sigma_X^2}}}{\sigma_X^2\sigma_Y^2}\left[\frac{\sqrt\pi erfc\left(\frac{w}{\sqrt\tau\sigma_X^2}\right)\exp\left(\frac{w^2}{\tau\sigma_X^4}\right)}{2\tau^\frac{3}{2}\sigma_X^2}\times
 \right. \\  \left.
\left\{\frac{\tau\sigma_X^4+2w^2}{2\tau\sigma_X^2}-w^2\right\}+\frac{w\ }{2\tau}\left(1-\frac{1}{\tau\sigma_X^2}\right)\right],   & \text{$w\geq0$}.
  \end{cases}
\end{equation}
where $\tau=\sfrac{(\sigma_X^2+\sigma_Y^2)}{(\sigma_X^2\sigma_Y^2)}$.

Thus, ABEP for $x_1$ is derived as
\begin{equation}
P_I^{(R_1D)}(e)= \sum_i p(i_{2nd}) \int_0^\infty Q\left(\sqrt{2}\zeta_i\right)p_{\zeta_i}(\zeta_i)d\zeta_i,
\end{equation}
where $\zeta_A\sim Z$ and $\zeta_B\sim W$ with $\sigma_X\triangleq\sigma_{R_1D}\sqrt{\sfrac{\beta_1P_R}{N_0}}$ and $\sigma_Y\triangleq\sigma_{R_2D}\sqrt{\sfrac{\beta_2P_R}{N_0}}$.

As in the downlink analysis at $R_2$, the BEP for $x_2$ symbols between $R_2$ and $D$ can be found by considering two cases of SIC (i.e., correct (Fig. 2.b.\textit{(ii)}) or erroneous detection (Fig. 2.b\textit{(iii)}) of $x_1$). Recalling  that symbols are affected by different channel fading coefficients, with the aid of (6)-(7), the BEP for $x_2$, in case correct detection of $x_1$, is given as
\begin{equation}
\begin{split}
&P_{II}^{(R_2D)}(e|_{\zeta_i\cap correct_{x_1}})=p(A_{2nd})\left(Q\left(\sqrt{2}\zeta_C\right) \right.\\
&\left.-Q\left(\sqrt{2}\zeta_A\right)\right)+p(B_{2nd}) Q\left(\sqrt{2}\zeta_C\right),\\
\end{split}
\end{equation}
and with the aid of (8)-(9), in case erroneous detection of $x_1$
\begin{equation}
\begin{split}
&P_{II}^{(R_2D)}(e|_{\zeta_i\cap error_{x_1}})=p(A_{2nd})Q\left(\sqrt{2}\zeta_D\right) \\
&+p(B_{2nd}) \left(Q\left(\sqrt{2}\zeta_B\right)- Q\left(\sqrt{2}\zeta_E\right) \right),\\
\end{split}
\end{equation}
where $\zeta_C=\sqrt{\sfrac{\beta_2P_R}{N_0}}\left|h_{R_2D}\right|$, $\zeta_D=\sfrac{\left(2\sqrt{\beta_1P_R}\left|h_{R_1D}\right|+\sqrt{\beta_2P_R}\left|h_{R_2D}\right|\right)}{\sqrt{N_0}}$ and $\zeta_E=\sfrac{\left(2\sqrt{\beta_1P_R}\left|h_{R_1D}\right|-\sqrt{\beta_2P_R}\left|h_{R_2D}\right|\right)}{\sqrt{N_0}}$.
The ABEP for $x_2$ in the second phase is obtained by averaging instantaneous channel coefficients after summing two cases (15) and (16) as in (17) (see bottom of the page), where $\zeta_D\sim Z$ and $\zeta_E\sim W$ with $\sigma_X\triangleq2\sigma_{R_1D}\sqrt{\sfrac{\beta_1P_R}{N_0}}$ and $\sigma_Y\triangleq\sigma_{R_2D}\sqrt{\sfrac{\beta_2P_R}{N_0}}$.
\begin{figure*}[b]
\hrulefill
\begin{equation}
\begin{split}
&P_{II}^{(R_2D)}(e)= 0.5(1-\sqrt{\frac{\beta_2 P_R\overline{\gamma}_{R_2D}}{1+\beta_2 P_R\overline{\gamma}_{R_2D}}})+p(A_{2nd})\left(-\int_0^\infty Q\left(\sqrt{2}\zeta_A\right)p_{\zeta_A}(\zeta_A)d\zeta_A+\int_0^\infty Q\left(\sqrt{2}\zeta_D\right)p_{\zeta_D}(\zeta_D)d\zeta_D\right)\\&+p(B_{2nd}) \left(\int_0^\infty Q\left(\sqrt{2}\zeta_B\right)p_{\zeta_B}(\zeta_B)d\zeta_B- \int_0^\infty Q\left(\sqrt{2}\zeta_E\right)p_{\zeta_E}(\zeta_E)d\zeta_E\right)
\end{split}
\end{equation}
\end{figure*}
\newtheorem{corollary}{\textbf{Corollary}}
\newtheorem{lemma}{\textbf{Lemma}}
\begin{corollary}
\textit{The priori probabilities of the superimposed symbols in the second phase have dominant effect on the error performance. Hence, we should obtain the priori probabilities to derive exact expressions.}
\end{corollary}
\begin{lemma}
\textit{It is mostly expected that the priori probabilities of events A and B have equal probability in the first phase. However, in the second phase of communication, $p(A_{2nd})\geq p(B_{2nd})$ (non-equiprobable) even if $p(A_{1st})=p(B_{1st})$ in the first phase. In particular, it is in the low SNR regime, hence the error performance of second phase is dominated by the symbols which belong to event A. We provide priori probabilities for two different channel realizations and power allocations in Table I.}
\begin{IEEEproof}
See Appendix.
\end{IEEEproof}
\begin{table}[!ht]
\centering
\caption{Priori probability of $p(A_{2nd})$}
\label{table1}
\begin{tabular}{|c|c|c|c|c|c|c|c|}
\hline
\multicolumn{4}{|c|}{Scenario I}&\multicolumn{4}{|c|}{Scenario II} \\ \hline
\multicolumn{4}{|c|}{$\alpha_1=0.8$, $\sigma_{SR_1}^2=1$, $\sigma_{SR_2}^2=2$}&\multicolumn{4}{c|}{$\alpha_1=0.7$, $\sigma_{SR_1}^2=2$, $\sigma_{SR_2}^2=10$} \\
\hline
\multicolumn{4}{|c|}{$SNR=\sfrac{P_S}{N_0}(dB)$} & \multicolumn{4}{c|}{$SNR=\sfrac{P_S}{N_0}(dB)$} \\
\hline
0&5&10&15&0&5&10&15 \\
\hline
.6021&.5949&.5559&.5223&.6415&.6100&.5588&.5237 \\
\hline
\end{tabular}
\end{table}
\end{lemma}
\subsubsection{Approximate Expressions for Second Phase (Uplink-NOMA)}
To the best of authors' knowledge, closed-form expressions cannot be derived for (14) and (17). Nevertheless, we provide a tight approximation for error analysis of second phase.
\begin{corollary}
\textit{Considering the values of $Q(.)$ function, we can easily see that $Q(\eta)>>Q(\xi)$ when $\xi>>\eta>0$. Hence, we can omit  $Q(\sqrt{2}\zeta_A)$ in (11) and (15), and  $Q(\sqrt{2}\zeta_D)$ in (16) since $\zeta_A>>\zeta_B$ and $\zeta_D>>\zeta_E$.}
\end{corollary}
\begin{lemma}
\textit{In case $X$ and $Y$ are Rayleigh distributed random variables and $W=X-Y$, it can be approximated that
\begin{equation*}
\int_0^\infty Q\left(W\right)p_{W}(w)dw\approx\frac{\sigma_Y^2}{\sigma_X^2+\sigma_Y^2}.
\end{equation*}}
\end{lemma}
\begin{IEEEproof}
See \cite[Section 3.3]{Kara2018d}.
\end{IEEEproof}
With the aid of Corollary 2 and Lemma 2, approximated BEP for second phase is obtained as
\begin{equation}
P_I^{(R_1D)}(e)= p(B_{2nd}) \frac{\beta_2P_R\sigma_{R_2D}^2}{\beta_1P_R\sigma_{R_1D}^2+\beta_2P_R\sigma_{R_2D}^2},
\end{equation}
and
\begin{equation}
\begin{split}
&P_{II}^{(R_2D)}(e)= 0.5(1-\sqrt{\frac{\beta_2 P_R\overline{\gamma}_{R_2D}}{1+\beta_2 P_R\overline{\gamma}_{R_2D}}}) +p(B_{2nd}) \\
&\left(\frac{\beta_2P_R\sigma_{R_2D}^2}{\beta_1P_R\sigma_{R_1D}^2+\beta_2P_R\sigma_{R_2D}^2} -\frac{\beta_2P_R\sigma_{R_2D}^2}{4\beta_1P_R\sigma_{R_1D}^2+\beta_2P_R\sigma_{R_2D}^2}\right).
\end{split}
\end{equation}

\subsection{End-to-End Error Analysis}
The e2e ABEP for NOMA-DRN is determined by averaging e2e ABEP of two symbols and it is given as
\begin{equation}
P^{(e2e)}(e)=\frac{P_I^{(e2e)}(e)+P_{II}^{(e2e)}(e)}{2},
\end{equation}
where $P_I^{(e2e)}(e)$ and $P_{II}^{(e2e)}(e)$ denote e2e ABEP of $x_1$ and $x_2$ symbols, respectively. With the law of total probability, they are derived as
\begin{equation}
\begin{split}
P_I^{(e2e)}(e)=&P_{I}^{(SR_1)}(e) (1-P_{I}^{(R_1D)}(e))\\
&+(1-P_{I}^{(SR_1)}(e))P_{I}^{(R_1D)}(e),
\end{split}
\end{equation}
and
\begin{equation}
\begin{split}
P_{II}^{(e2e)}(e)=&P_{II}^{(SR_2)}(e) (1-P_{II}^{(R_2D)}(e))\\
&+(1-P_{II}^{(SR_2)}(e))P_{II}^{(R_2D)}(e).
\end{split}
\end{equation}
\subsection{High-SNR (Error Floor) Analysis}
NOMA-DRN has an interference-limited error performance and it is dominated by the second phase of communication (i.e., uplink-NOMA). Thus, its performance has an error floor even if it is in the very high-SNR regime. In order to derive this error floor, we provide error analysis when transmit SNR ( $\sfrac{P_S}{N_0}$ for source and $\sfrac{P_R}{N_0}$ for relays) are very high (i.e., $SNR\rightarrow\infty$).
According to provided ABEPs of the first phase in (5) and (10), one can easily see that $\lim_{SNR\rightarrow\infty}{P_{I}^{(SR_2)}(e)}=0$ and $\lim_{SNR\rightarrow\infty}{P_{II}^{(SR_2)}(e)}=0$. Thus, the e2e ABEPs of symbols given in (21) and (22) turn out to be
\begin{subequations}
\begin{equation}
\lim_{SNR\rightarrow\infty}{P_I^{(e2e)}(e)}=\lim_{SNR\rightarrow\infty}{P_I^{(R_1D)}(e)},
\end{equation}
\begin{equation}
\lim_{SNR\rightarrow\infty}{P_{II}^{(e2e)}(e)}=\lim_{SNR\rightarrow\infty}{P_{II}^{(R_2D)}(e)}.
\end{equation}
\end{subequations}
According to provided priori probability analysis between (A.1) and (A.4), we can easily see that $\lim_{SNR\rightarrow\infty}{p(B_{2nd}) }=\sfrac{1}{2}$. Therefore, by adopting ABEPs given in (18) and (19), error floors for high SNR are determined as
\begin{subequations}
\begin{equation}
 \lim_{SNR\rightarrow\infty}{P_I^{(e2e)}(e)}=\frac{1}{2}\frac{\beta_2\sigma_{R_2D}^2}{\beta_1\sigma_{R_1D}^2+\beta_2\sigma_{R_2D}^2},
\end{equation}
\begin{equation}
\begin{split}
&\lim_{SNR\rightarrow\infty}{P_{II}^{(e2e)}(e)}= \\
&\frac{1}{2}\left(\frac{\beta_2\sigma_{R_2D}^2}{\beta_1\sigma_{R_1D}^2+\beta_2\sigma_{R_2D}^2}-\frac{\beta_2\sigma_{R_2D}^2}{4\beta_1\sigma_{R_1D}^2+\beta_2\sigma_{R_2D}^2}\right),
\end{split}
\end{equation}
\begin{equation}
\begin{split}
&\lim_{SNR\rightarrow\infty}{P^{(e2e)}(e)}= \\
&\frac{1}{2}\left(\frac{\beta_2\sigma_{R_2D}^2}{\beta_1\sigma_{R_1D}^2+\beta_2\sigma_{R_2D}^2}-\frac{1}{2}\frac{\beta_2\sigma_{R_2D}^2}{4\beta_1\sigma_{R_1D}^2+\beta_2\sigma_{R_2D}^2}\right).
\end{split}
\end{equation}
\end{subequations}
\begin{figure}[!t]
    \centering
    \includegraphics[width=8cm,height=5.4cm]{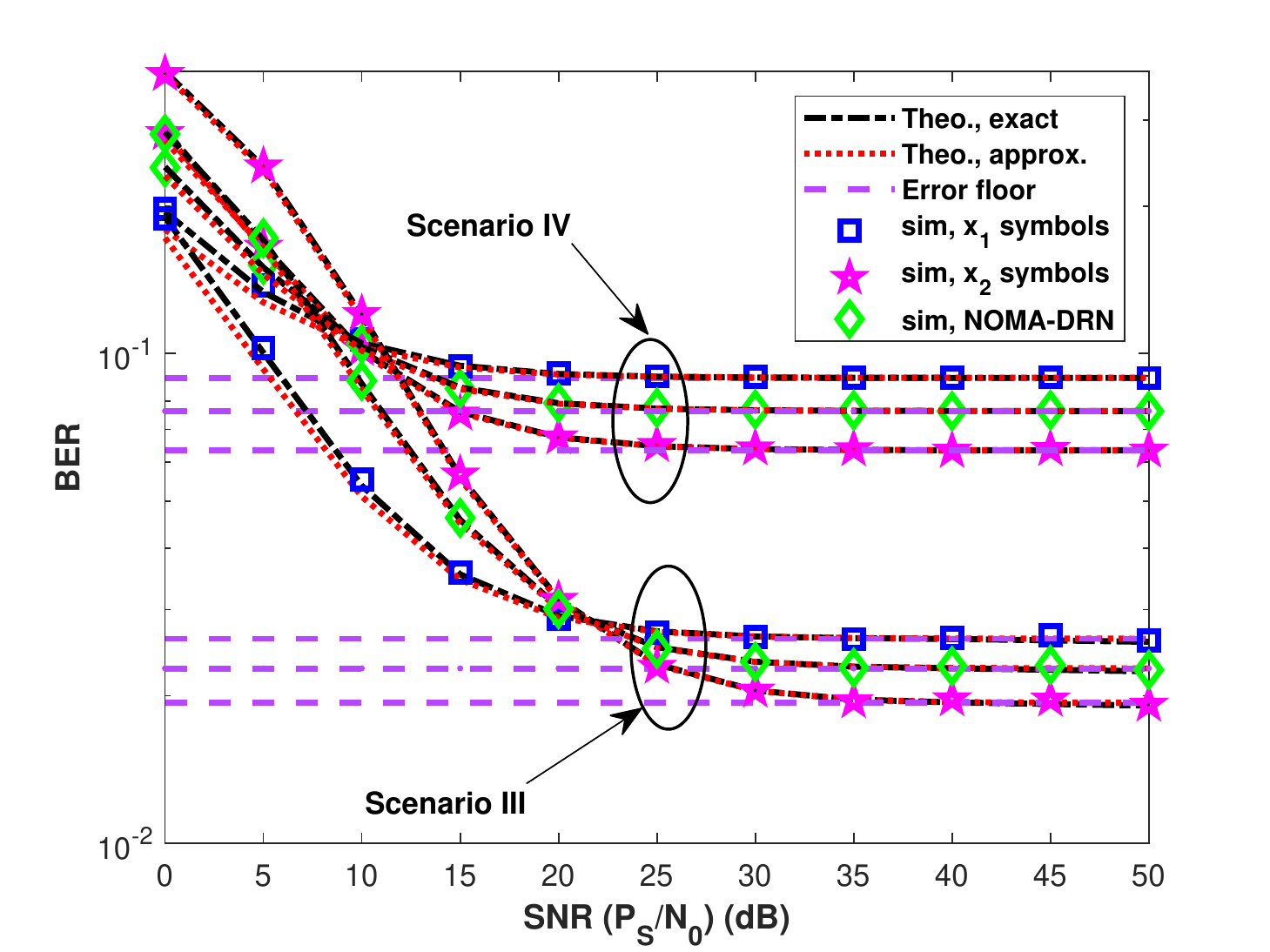}
    \caption{Error performance of NOMA-DRN}
    \label{fig2}
\end{figure}
\section{Numerical Results}
In Fig. 3, we present error performance of NOMA-DRN. In Scenario III, $\sigma_{SR_1}^2=1$, $\sigma_{SR_2}^2=10$, $\sigma_{R_1D}^2=9$, $\sigma_{R_2D}^2=2$ and, Scenario IV, $\sigma_{SR_1}^2=2$, $\sigma_{SR_2}^2=10$, $\sigma_{R_1D}^2=9$, $\sigma_{R_2D}^2=3$. In all simulations, $P_S=P_R$ is assumed. The power allocation coefficients are given $\alpha_1=.9602$, $\beta_1=.8011$ and $\alpha_1=.8816$, $\beta_1=.6055$ which are given as sub-optimum coefficients for given scenarios, respectively \cite{Wan2019}. In the results, we provide e2e bit error rate (BER) for $x_1$, $x_2$ symbols and average BER for NOMA-DRN. Based on the simulations, it is noteworthy that derived one-degree integral form of exact ABEP matches perfectly with simulation. In addition, provided approximate expression match well also. Furthermore, as it is proved in the previous section, NOMA-DRN has an error floor and its error performance is limited even though transmit SNR is increased. This is dominated by the poor error performance of second phase (uplink-NOMA) and derived error floor expressions are perfectly match in high-SNR regime. It is also proved that NOMA-DRN has a non-equiprobable communication in the second phase and derived priori probabilities are correct.

\begin{figure}[!t]
    \centering
    \includegraphics[width=8cm,height=6cm]{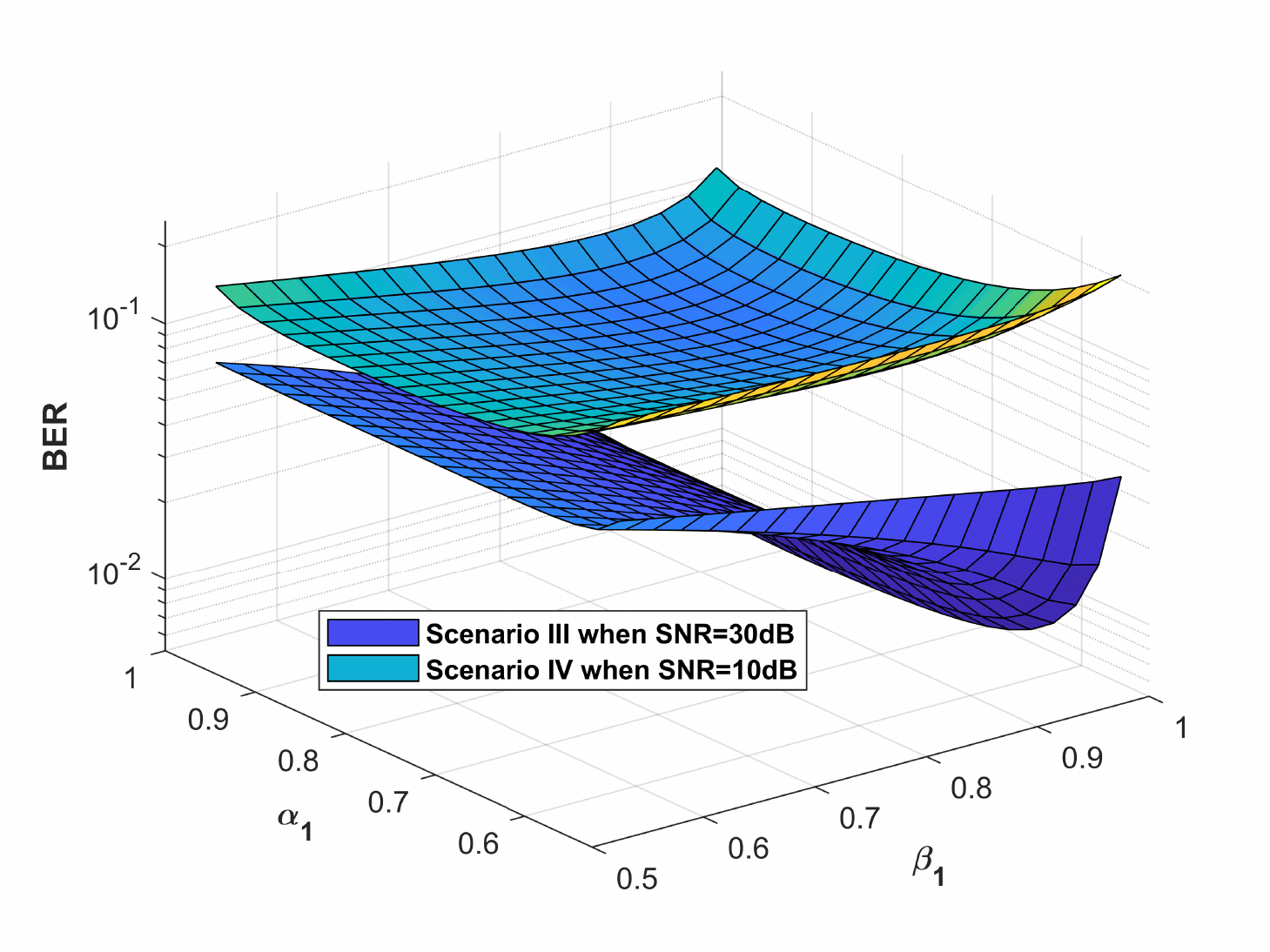}
    \caption{BER of NOMA-DRN with the change of power allocation pairs}
    \label{fig3}
\end{figure}
In order to show the effect of power allocation coefficients (i.e., $\alpha_k$, $\beta_k$, $k=1,2$), we provide BER of NOMA-DRN with the change of $\alpha_1$ and $\beta_1$, in Fig. 4. E2e BER is presented for Scenario III when $SNR=30dB$ and for Scenario IV when $SNR=10dB$, respectively. Based on provided simulation results, optimum power allocation pairs which minimize e2e BER of NOMA-DRN for given scenarios are $\alpha_1 \simeq.875$, $\beta_1 \simeq.975$, and  $\alpha_1 \simeq.875$, $\beta_1 \simeq.875$, respectively. Nevertheless, it is noteworthy that NOMA-DRN will still have an error floor even if optimum pairs are used since it is caused by second phase of communication (uplink-NOMA).

NOMA-DRN has a non-equiprobable communication and these priori probabilities of symbols have dominant effect on the error performance of second phase of NOMA-DRN. In order to validate derived priori probabilities (i.e., $p(A_{2nd})$, $p(B_{2nd})$) and to emphasize their effect, we present error performance for second phase of NOMA-DRN in Fig. 5 for the channel realizations given in Table I. Additionally, in Scenario I, $\beta_1=0.8$, $\sigma_{R_1D}^2=2$, $\sigma_{R_2D}^2=1$ and in Scenario II, $\beta_1=0.7$, $\sigma_{R_1D}^2=10$, $\sigma_{R_2D}^2=2$ are assumed. We provide theoretical curves for non-equiprobable communication according to derived priori probabilities in Lemma 1. We provide theoretical curves for  also equiprobable communication as if $p(A_{2nd})=p(B_{2nd})$. One can easily see that derived priori probabilities are correct and analytical analysis match perfectly with simulations. The effect of this non-equiprobable communication has dominant effect especially in the low SNR regime (green circle in Fig. 5). Based on simulation results, in the low SNR regime, the error performance is not always improved by the increase in SNR unlike expected. This can be explained as follows: In that region, $p(A_{2nd})>p(B_{2nd})$ and the BEP for the symbols in event A is lower than the event B since the symbols in event A have higher energy levels (see Fig. 2). Thus, BER performance is better in that region. However, with the increase in SNR, the priori probabilities of A and B become closer and the error performance is dominated by the event B since the symbols in event B have much less energy levels. The increase in SNR cannot overcome the penalty of the domination of event B, hence the error performance gets worse.
\begin{figure}[!t]
    \centering
    \includegraphics[width=8cm,height=5.4cm]{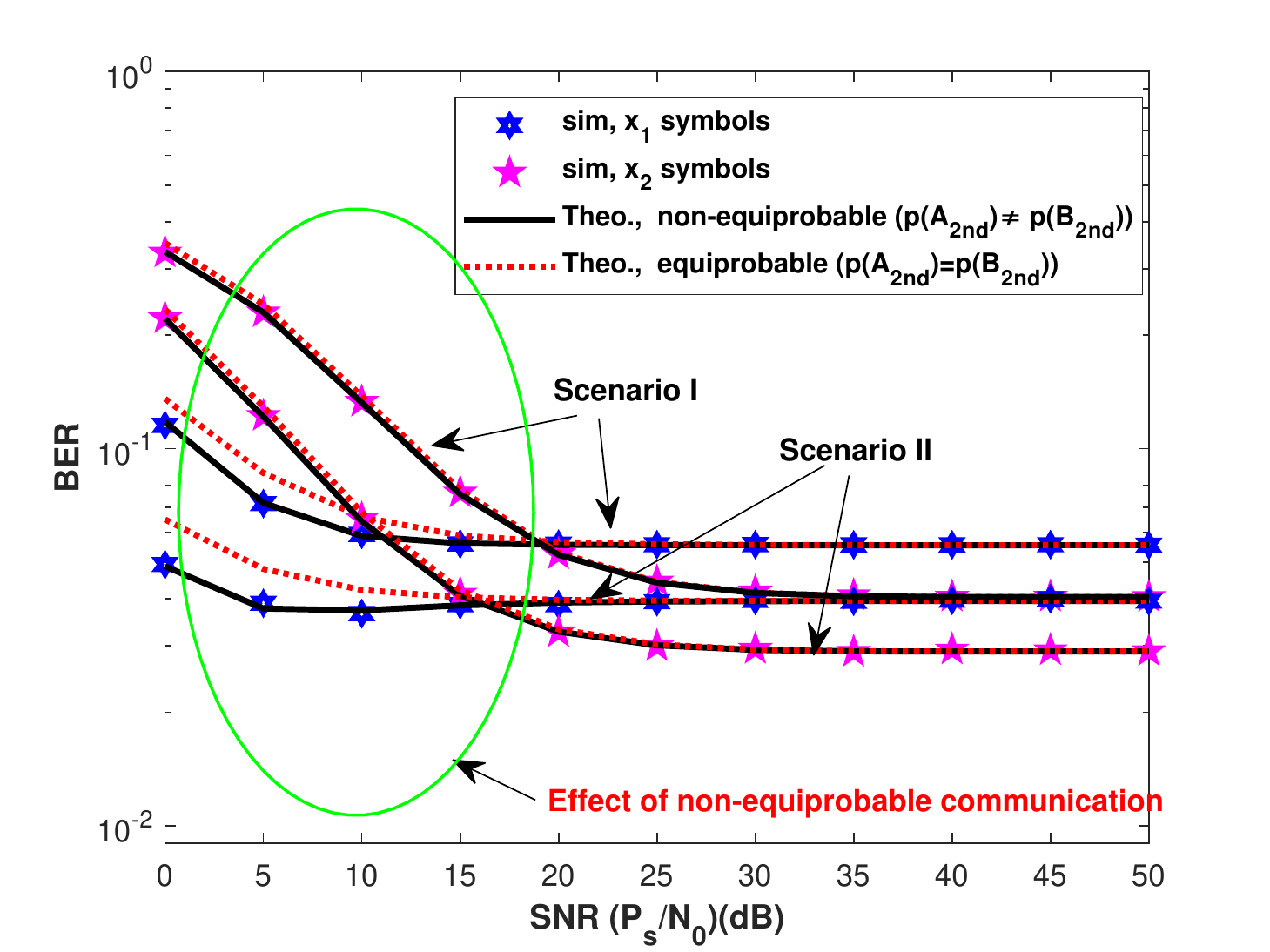}
    \caption{Error performance for the second phase of NOMA-DRN}
    \label{fig4}
\end{figure}
\section{Conclusion}
In this paper, we analyze error performance of NOMA-DRN and derive exact BEP in a single integral form. Nevertheless, we provide a tight approximate BEP in a closed-form and an error floor analysis for high-SNR regime. We prove that NOMA-DRN includes non-equiprobable symbols in the second phase of communication and we derive these priori probabilities. All derived expressions are validated via computer simulations. In this paper, the analysis is conducted for BPSK. The extension of analysis for higher modulations is very complex since each combinations of symbols will have different energy levels after superposition coding at BS or superimposed at destination and all these combinations should be considered. Furthermore, as it is proved in this paper, non-equiprobable communication should be also considered for all symbols. Nevertheless, this analysis can be handled by utilizing symbol error rate analysis of conventional NOMA for M-QAM as given in \cite{Lee2019}. This is seen as a future work. It is shown that NOMA-DRN has a poor error performance although NOMA-CRS are very promising in terms of achievable rate. Hence, the error performance should be also considered and theoretical error analysis should be handled for other NOMA-CRS schemes. Lastly, we believe that DRN has a huge potential for distributed coded schemes, thus channel coding techniques such as Turbo coding, LPDC or Polar coding can be implemented in NOMA-DRN to enhance error performance \cite{Ozturk2015}. These are seen as directions of future works.
\appendix
\renewcommand{\theequation}{\thesection.\arabic{equation}}
\setcounter{equation}{0}
We consider the probability of the event A which may occur in the second phase of communication. One can easily see that this event depends on the detections at the relays and the priori probability of events in the first phase. Firstly, we assume that the event A has already occurred in the first phase. In this case, the event A occurs in the second phase only if both relays detect correctly or erroneously their related symbols ($R_1\rightarrow x_1$ and $R_2\rightarrow x_2$) at the same time, and the conditional probability is given as
\begin{equation}
\begin{split}
&p(A_{2nd}|_{A_{1st}})=(1-P_I^{(SR_1)}(e|_{A_{1st}}))(1-P_{II}^{(SR_2)}(e|_{A_{1st}}))\\
&+P_I^{(SR_1)}(e|_{A_{1st}})P_{II}^{(SR_2)}(e|_{A_{1st}}).
\end{split}
\end{equation}
On the other hand, if B occurs in the first phase, the event A occurs in the second phase only if one of the relays detects erroneously the related symbols. It is given as
\begin{equation}
\begin{split}
&p(A_{2nd}|_{B_{1st}})=P_I^{(SR_1)}(e|_{B_{1st}})(1-P_{II}^{(SR_2)}(e|_{B_{1st}}))\\
&+(1-P_I^{(SR_1)}(e|_{B_{1st}}))P_{II}^{(SR_2)}(e|_{B_{1st}}),
\end{split}
\end{equation}
where conditional probabilities denote the error probability when the related condition event occurred. Total probability is given as
\begin{equation}
p(A_{2nd})=p(A_{1st})p(A_{2nd}|_{A_{1st}})+p(B_{1st})p(A_{2nd}|_{B_{1st}}),
\end{equation}
and with the aid of total probability law
\begin{equation}
p(B_{2nd})=1-p(A_{2nd}).
\end{equation}
Substituting related expressions from (5) and (10) into (A.3) and (A.4), we derive the priori probabilities. So the proof is completed.

%



\ifCLASSOPTIONcaptionsoff
  \newpage
\fi



%
\bibliographystyle{IEEEtran}
\bibliography{IEEEabrv,noma_drn}

%






\end{document}